\documentclass[journal]{IEEEtran}

\usepackage[dvips]{color}
\usepackage{graphicx}
\usepackage{amsmath}
\begin{document}
%
% paper title
% can use linebreaks \\ within to get better formatting as desired
\title{Security versus Reliability Analysis of Opportunistic Relaying}
%
%
% author names and IEEE memberships
% note positions of commas and nonbreaking spaces ( ~ ) LaTeX will not break
% a structure at a ~ so this keeps an author's name from being broken across
% two lines.
% use \thanks{} to gain access to the first footnote area
% a separate \thanks must be used for each paragraph as LaTeX2e's \thanks
% was not built to handle multiple paragraphs
%

\markboth{IEEE Transactions on Vehicular Technology (accepted to appear)}%
{Yulong Zou \MakeLowercase{\textit{et al.}}: Security versus Reliability Analysis of Opportunistic Relaying}

\author{Yulong~Zou,~\IEEEmembership{Senior Member,~IEEE,}
        Xianbin~Wang,~\IEEEmembership{Senior Member,~IEEE,}
        Weiming~Shen,~\IEEEmembership{Fellow,~IEEE,}
        and Lajos Hanzo,~\IEEEmembership{Fellow,~IEEE}

\thanks{Copyright (c) 2013 IEEE. Personal use of this material is permitted. However, permission to use this material for any other purposes must be obtained from the IEEE by sending a request to pubs-permissions@ieee.org.}
\thanks{Manuscript received July 24, 2013; revised October 15,
2013; accepted November 22, 2013. The editor coordinating the review of this paper and approving it for publication was Prof. Hsiao-Hwa Chen.}
\thanks{This work was partially supported by the National Natural Science Foundation of China (Grant Nos. 61302104, 61271240), the Scientific Research Foundation of Nanjing University of Posts and Telecommunications (Grant No. NY213014), and the Auto21 Network of Centre of Excellence, Canada.}
\thanks{Y. Zou is with the School of Telecommunications and Information Engineering, Nanjing University of Posts and Telecommunications, Nanjing, Jiangsu 210003, P. R. China. (E-mail:
yulong.zou@njupt.edu.cn).}
\thanks{X. Wang is with the Electrical and Computer Engineering Department, Western University, London, Ontario, Canada, N6A 5B9. (E-mail: xianbin.wang@uwo.ca).}
\thanks{W. Shen is with the College of Electronics and Information Engineering, Tongji University, Shanghai, China, and with National Research Council, Ottawa, Canada. (E-mail: weiming.shen@nrc.gc.ca).}
\thanks{L. Hanzo is with the Department of Electronics and Computer Science, University of
Southampton, Southampton, United Kingdom. (E-mail: lh@ecs.soton.ac.uk)}

}

%on the Reliable and Secure Sensor Networks for Factory Automation

% make the title area
\maketitle

\begin{abstract}
Physical-layer security is emerging as a promising paradigm of securing wireless communications against eavesdropping between legitimate users, when the main link spanning from source to
destination has better propagation conditions than the wiretap link
from source to eavesdropper. In this paper, we identify and analyze
the tradeoffs between the security and reliability of wireless
communications in the presence of eavesdropping attacks. Typically,
the reliability of the main link can be improved by increasing the
source's transmit power (or decreasing its date rate) to reduce the
outage probability, which unfortunately increases the risk that an
eavesdropper succeeds in intercepting the source message through the
wiretap link, since the outage probability of the wiretap link also
decreases when a higher transmit power (or lower date rate) is used.
We characterize the security-reliability tradeoffs (SRT) of
conventional direct transmission from source to destination in the
presence of an eavesdropper, where the security and reliability are
quantified in terms of the intercept probability by an eavesdropper
and the outage probability experienced at the destination,
respectively. In order to improve the SRT, we then propose
opportunistic relay selection (ORS) and quantify the attainable SRT
improvement upon increasing the number of relays. It is shown that given the maximum tolerable intercept probability, the outage probability of our ORS scheme approaches zero for $N \to \infty$, where $N$ is the number of relays. Conversely, given the maximum tolerable outage probability, the intercept probability of our ORS scheme tends to zero for $N \to \infty$.

\end{abstract}

\begin{IEEEkeywords}
Security-reliability tradeoff, physical-layer security,
opportunistic relay selection, intercept probability, outage
probability, cooperative communications.

\end{IEEEkeywords}

\IEEEpeerreviewmaketitle

\section{Introduction}
% The very first letter is a 2 line initial drop letter followed
% by the rest of the first word in caps.
%
% form to use if the first word consists of a single letter:
% \IEEEPARstart{A}{demo} file is ....
%
% form to use if you need the single drop letter followed by
% normal text (unknown if ever used by IEEE):
% \IEEEPARstart{A}{}demo file is ....
%
% Some journals put the first two words in caps:
% \IEEEPARstart{T}{his demo} file is ....
%
% Here we have the typical use of a "T" for an initial drop letter
% and "HIS" in caps to complete the first word.

\IEEEPARstart {A}{t} the time of writing, the Internet is typically
accessed through the wireless infrastructure (e.g., cellular
networks and Wi-Fi) [1]. Consequently, the security of wireless
communications plays an increasingly important role in the
cybercrime defense against unauthorized activities. Moreover, due to
the broadcast nature of the wireless medium, transmissions between
legitimate users may readily be overheard and intercepted by
unauthorized parties, which makes wireless transmission vulnerable
to potential eavesdropping attacks. As a result, wireless security
has received growing research attention in recent years. In existing
wireless communication systems, cryptographic techniques are used
for preventing an unauthorized eavesdropper from intercepting
message transmissions between legitimate users [2], [3]. Although
the cryptographic methods indeed improve the communication security,
this comes at the expense of increased communication and
computational overheads. To be specific, the increased complexity of
encryption algorithm enhances the security level of wireless
communications, which unfortunately requires more processing
resources for encryption as well as decryption and increases latency
imposed. Furthermore, the encryption introduces additional
redundancy and hence results in an increased overhead. Additionally,
the encryption may still be decrypted by an eavesdropper using an
exhaustive key search (also known as brute-force attack).

As an alternative, physical-layer security (PLS) is emerging as a
promising secure wireless communications paradigm relying on
exploiting the physical characteristics of wireless channels for
protection against eavesdropping attacks. The root of PLS may be
traced back to the 1970s [4], where a discrete memoryless wiretap
channel (WTC) consisting of a single source, destination and
eavesdropper is investigated in an information-theoretic sense. It was shown that reliable data transmission at non-zero rates may be
achieved in perfect secrecy.  In [5], the authors extended Wyner's
results originally derived for discrete memoryless WTCs to Gaussian
WTCs and quantified the \emph{secrecy capacity} (SC), namely the
difference between the channel capacity of the main link (from source to destination) and of the WTC (from source to eavesdropper).
In [6], the impact of feedback on WTCs was further investigated in
terms of their SC, showing that reliable and secure transmission is
still possible, even when the main link is inferior to the wiretap
link by exploiting the feedback information. However, the SC of
wireless transmission is severely degraded by the time-varying
multipath fading effects.

To mitigate the time-varying fading, considerable efforts have been
invested in improving the SC of wireless transmissions by exploiting
multiple antennas, since they enhance the channel capacity [7],
[8]. In [9], the authors studied the SC of multiple-input
single-output (MISO) WTCs. By contrast, the multiple-input
multiple-output (MIMO) WTC was examined in [10], where the source (S), destination (D) and eavesdropper (E) rely on multiple antennas and the SC is characterized by using a generalized-singular-value-decomposition. In [11], the authors
investigated the MIMO broadcast WTC and proved that the secrecy
capacity of MIMO broadcast channels is given by the difference between the capacities of the main and wiretap links. As a later advance, user cooperation [12], [13] was interpreted as a virtual MIMO formed by the cooperating single-antenna users for achieving spatial diversity gain and hence improves the SC of WTCs [14]-[16].  In [15], cooperative jamming was advocated for improving the PLS, where multiple users cooperate with each other by forming a coalition against eavesdropping. In [16], cooperative decode-and-forward (DF) and {amplify-and-forward (AF)} schemes were conceived and it was shown that the SC can be significantly improved by user cooperation.

In the PLS literature, E is routinely assumed to be aware of all
system parameters of the legitimate link between S and D,
including the carrier frequency, bandwidth, coding and modulation
scheme, encryption algorithm and secrecy key, etc. Since wireless
systems are standardized, the aforementioned operating parameters
(e.g., carrier frequency, bandwidth, etc.) may be readily inferred by exploiting the weaknesses of the protocols.  In the PLS approaches of [4]-[6], the encoder of the source is designed to maximize the data
rate $R_d$ at D and the equivocation rate $R_e$ at E. As discussed in [17] and [18], the idealized or perfect secrecy rate $R_d$ is
achieved, if the data rate is as high as the equivocation rate (i.e., we have $R_d=R_e$), which implies that the data rate $R_d$ is
achievable at D, while the mutual information between the transmission of S and E becomes zero.  Moreover, the SC is the
highest achievable perfectly secure rate and hence, provided that the data rate is set below the SC, reliable transmissions between S and D may be achieved in perfect secrecy. However, in practice, it is challenging to devise an ideal scheme for ensuring that D can reliably communicate with S at a non-zero rate (less than the SC) and, at the same time, E fails to decode the {source-signal (SS)}.

As a result, we do not rely on the above-mentioned idealized scheme
routinely used in the PLS literature~[4] - [6], [9] - [11] and
operating exactly at the SC, but rather at a lower rate $R_d$ for
maintaining PLS. As discussed in [19], a low intercept probability
(IP) can achieved by constraining the capacity of E. For example, when the WTC capacity is lower than the main channel's capacity, S may adjust its data rate to be between the capacity of the main and that of the WTCs for depriving E from achieving an arbitrarily low decoding error rate, while ensuring reliable communication for D. More specifically, according to Shannon~[20], when the capacity of the WTC spanning from S to E is lower than the data rate, E fails to decode the SS, while the legitimate transmission remains secure. However, if the WTC capacity becomes higher than the data rate of D, E may succeed in decoding the SS and hence an intercept event occurs. Although increasing the data rate (or decreasing the transmit power of S) may reduce the IP and improves the level of security, this comes at the cost of transmission reliability degradation, since the outage probability (OP) of the main link also increases, when a higher data rate (or lower transmit power) is used at S. Therefore, our motivation is to strike a security versus reliability tradeoff
(SRT). Although the notion of SRT was studied in [21], this contribution was mainly focused on the employment of various block cipher encryption algorithms to defend against eavesdropping attacks. By contrast, in our work PLS - rather than block ciphering - is invoked for characterizing the SRT performance attained in fading wireless environments.

The main contributions of this paper are summarized as follows.
Firstly, we characterize the SRT in terms of the probability that E
succeeds in intercepting the SS versus the probability that an outage event occurs at D, respectively. Secondly, we quantify the benefits of opportunistic relay selection (ORS) in terms of improving the SRT, especially upon increasing the number of relays, while ensuring that the best relay is activated for reducing both the IP and OP.

The remainder of this paper is organized as follows. Section II
describes our system model while in Section III we propose an ORS
scheme and carry out its SRT analysis in Rayleigh fading channels. In Section IV, we provide numerical SRT results and show that the ORS
scheme always outperforms the conventional direct transmission,
especially as the number of relays increases. Finally, Section V
presents our concluding remarks.

\section{PLS for Wireless Communications}
\subsection{System Model}
\begin{figure}
  \centering
  {\includegraphics[scale=0.65]{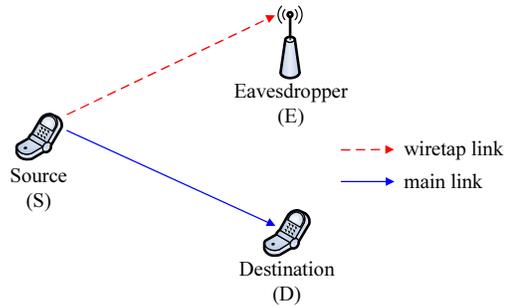}\\
  \caption{A wireless network consists of one source (S) and one destination (D) in the presence
  of an eavesdropper (E).}\label{Fig1}}
\end{figure}

Let us now present our system model and analyze the SRT in Rayleigh
fading channels. Consider the wireless scenario of Fig. 1 consisting
of a single S, D and E, where the solid and dashed lines represent the S-D main link and S-E WTC, respectively. {Observe that the system model of Fig. 1 is applicable to diverse practical wireless systems,
including the family of wireless local area networks (WLANs), wireless sensor networks (WSNs), cellular networks, mobile ad hoc networks (MANETs) and so on.} In Fig. 1, S is characterized by its transmit power $P$ and data rate $R_d$ given by the Shannon-capacity. This is in contrast to the existing trends in the PLS literature [4]-[6] and [9]-[11], where typically an ideal scheme operating exactly at the SC limit is assumed so that D can reliably communicate with S, while E fails to decode the SS. When S transmits its signal $x$ at a power $P$ and rate $R_d$, E may overhear the transmission of S and attempts to decode the SS $x$. If E succeeds in decoding $x$, an intercept event occurs. When S transmits $x$ at a power $P$ and rate $R_d$, we may express the signal received at D as
\begin{equation}\label{equa1} y_d=\sqrt{P}h_{sd}x+n_d,
\end{equation}
where $h_{sd}$ represents the fading coefficient of the S-D channel
and $n_d $ is the zero-mean additive white Gaussian noise (AWGN) of
variance $N_0$. Again, in line with~[9]-[11], E is assumed to know all the parameters of S. Hence, the signal received by E is written as
\begin{equation}\label{equa2} y_e=\sqrt{P}h_{se}x+n_e,
\end{equation}
where $h_{se}$ represents the fading coefficient of the S-E channel
and $n_e$ is also a zero-mean AWGN process with a variance of $N_0$.
According to Shannon [20], the S-D link's capacity is
\begin{equation}\label{equa3}
C_{sd}=\log_2\left(1+{|h_{sd}|^2\gamma}\right),
\end{equation}
where $\gamma=\frac{P}{N_0}$ is the signal-to-noise ratio (SNR).
Similarly, using (2), the capacity of the S-E WTC is
formulated as
\begin{equation}\label{equa4}
C_{se}=\log_2\left(1+{|h_{se}|^2\gamma}\right).
\end{equation}
Since both the main and the WTC are modeled as Rayleigh
fading channels, $|h_{sd}|^2$ and $|h_{se}|^2$ are
exponentially distributed random variables with means of
$\sigma_{sd}^2$ and $\sigma_{se}^2$, respectively.

\subsection{Security-Reliability Tradeoff Analysis}
An intercept event is encountered when the WTC capacity becomes
higher than the data rate, hence the IP ${P_{{\mathop{\rm int}} }}$
of direct transmission becomes
\begin{equation}\label{equa5}
{P_{{\mathop{\rm int}} }} = \Pr \left( {{C_{se}} > R_d} \right).
\end{equation}
Substituting $C_{se}$ from (4) into (5) gives the IP
\begin{equation}\label{equa6}
\begin{split}
{P_{{\mathop{\rm int}} }} &= \Pr \left( {{{\log }_2}(1 + |{h_{se}}{|^2}\gamma ) > R_d} \right)\\
&= \Pr \left( {|{h_{se}}{|^2} > \alpha } \right),
\end{split}
\end{equation}
where $\alpha  = \frac{{{2^{R_d}} - 1}}{\gamma }$. Since
$|h_{se}|^2$ obeys the exponential distribution, the IP of (6)
becomes
\begin{equation}\label{equa7} {P_{{\mathop{\rm int}} }} = \exp ( -
\frac{\alpha }{{\sigma _{se}^2}}),
\end{equation}
where again, $\sigma _{se}^2 = E\left( {|{h_{se}}{|^2}} \right)$ is
the expected value of $|h_{se}|^2$. Additionally, according to
Shannon [20], the OP ${P_{{\mathop{\rm out}} }}$ of direct
transmission from S to D is obtained as
\begin{equation}\label{equa8}
{P_{{\mathop{\rm out}} }} = \Pr \left( {{C_{sd}} < R_d} \right),
\end{equation}
where $C_{sd}$ is given by (3). Substituting (3) into (8) yields
\begin{equation}\label{equa9}
\begin{split}
{P_{{\rm{out}}}} &= \Pr \left( {{{\log }_2}(1 + |{h_{sd}}{|^2}\gamma ) < R_d} \right)\\
%&= \Pr \left( {|{h_{sd}}{|^2} < \alpha } \right)\\
&= 1 - \exp ( - \frac{\alpha }{{\sigma _{sd}^2}}),
\end{split}
\end{equation}
where $\sigma _{sd}^2 = E\left( {|{h_{sd}}{|^2}} \right)$ is the
expected value of $|h_{sd}|^2$. Combining (7) and (9) yields
\begin{equation}\label{equa10} {P_{{\rm{out}}}} = 1 -
{({P_{{\mathop{\rm int}} }})^{\sigma _{se}^2/\sigma _{sd}^2}},
\end{equation}
where ${0 \le P_{{\mathop{\rm int}} }} \le 1$, $\sigma _{se}^2 > 0$,
and $\sigma _{sd}^2 > 0$. It is observed from (10) that increasing
$P_{\rm{int}}$ reduces $P_{\rm{out}}$, again indicating a tradeoff
between security and reliability, which essentially hinges on the
average channel gains $\sigma^2_{se}$ and $\sigma^2_{sd}$, but it is
independent of the transmit power $P$ and data rate $R_d$. Hence, the SRT cannot be improved by adjusting $P$ and $R_d$. This motivates the employment of ORS for the SRT improvements.

\section{Opportunistic Relay Selection in Cooperative Wireless Networks}

\subsection{ORS Scheme}
\begin{figure}
  \centering
  {\includegraphics[scale=0.65]{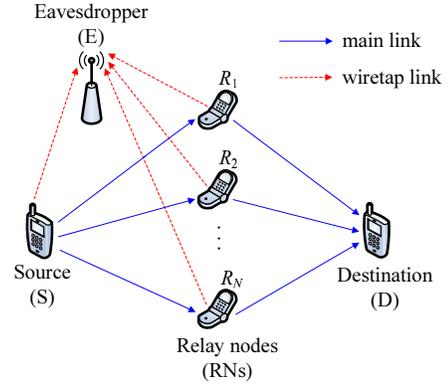}\\
  \caption{A cooperative wireless network consists of one source (S), one destination (D),
  and $N$ relay nodes (RNs) in the presence of an eavesdropper (E).}\label{Fig2}}
\end{figure}
In the cooperative wireless network of Fig. 2, $N$ RNs assist the
legitimate transmission from S to D, where the S-D direct link is
assumed to be unavailable (owing to its low quality). {At the time of writing, such a relay architecture has been adopted in commercial wireless networks such as for example the IEEE 802.16j/m or the long term evolution (LTE)-advanced cellular system, where relay stations may be introduced for assisting data transmissions between a base station and a user terminal.} An eavesdropper is located randomly around the S and RNs. We consider the worst-case scenario, where E overhears the transmissions of both the S and RNs and attempts to decode the SS. We denote the set of $N$ RNs by ${\cal{R}} = \{ {R_i}|i = 1,2, \cdots ,N\} $, where the DF protocol is employed.  Similar findings are valid for AF relaying protocol. During the ORS process, S first broadcasts its signal and the $N$ RNs attempt to decode it. Those RNs that succeed in perfectly decoding the source signal are represented by the successful decoding set ${\cal {D}}$. Given $N$ RNs, there are $2^N$ possible S-R pair combinations from the full set $\cal{R}$ of $N$ RNs, thus the resultant successful decoding set ${\cal {D}}$ is given by
\begin{equation}\label{equa11}
\Omega  = \left\{ {\emptyset ,{\cal {D}}_1 ,{\cal {D}}_2 , \cdots ,{\cal {D}}_n , \cdots ,{\cal {D}}_{2^N  - 1} } \right\},
\end{equation}
where $ \emptyset$ represents an empty set, while $ {\cal {D}}_n$ is a non-empty subset from $N$ RNs. If the decoding set is empty, since no RN succeeds in perfectly decoding the SS, all RNs will remain
silent. Thus, D is unable to infer the SS from the RNs. If the
successful decoding set ${\cal {D}}$ is non-empty, a specific RN will be opportunistically chosen from ${\cal {D}}$ for forwarding its
decoded signal to D.  When S transmits its signal $ x $ at a power $P$ and rate $R_d$, the channel capacities of the S-R$_{i}$ and S-E links relying on ORS are given by
%\begin{equation}\label{equa12}
%y_i=\sqrt{P}h_{si}x+n_i,
%\end{equation}
%where $h_{si}$ represents the fading coefficient of channel from
%source to relay $r_i$ and $n_i $ represents zero-mean additive white
%Gaussian noise (AWGN) with variance $N_0$. Meanwhile, due to the
%broadcast nature of wireless transmission, the eavesdropper can
%overhear the source transmission and the corresponding received
%signal is given by
%\begin{equation}\label{equa13}
%y_e=\sqrt{P}h_{se}x+n_e.
%\end{equation}
\begin{equation}\label{equa12}
C^{\textrm{ORS}}_{si}=\frac{1}{2}\log_2\left(1+{|h_{si}|^2\gamma}\right),
\end{equation}
and
\begin{equation}\label{equa13}
C^{\textrm{ORS}}_{se}=\frac{1}{2}\log_2\left(1+{|h_{se}|^2\gamma}\right),
\end{equation}
where the capacity is halved, because two orthogonal time slots are
required for completing the S-D transmission via R${_i}$. Again,
when $C^{\textrm{ORS}}_{si}<R_d$, the RN R$_i$ is unable to decode
the SS $x$. Thus, the scenario of ${\cal {D}} =\emptyset$ is
described as
\begin{equation}\label{equa14}
\frac{1}{2}{\log _2}\left( {1 +
|{h_{si}}{|^2}\gamma } \right) < R_d,\textrm{ }R_i \in {\cal{R}}.
\end{equation}
Similarly, the event ${\cal {D}} ={\cal {D}}_n$ is formulated as
\begin{equation}\label{equa15}
\begin{split}
&\frac{1}{2}{\log _2}\left( {1 + |{h_{si}}{|^2}\gamma } \right) > R_d,\textrm{ } R_i \in {{\cal {D}}_n}\\
&\frac{1}{2}{\log _2}\left( {1 + |{h_{sj}}{|^2}\gamma } \right) < R_d,\textrm{ } R_j \in {{\bar {\cal D}}_n},
\end{split}
\end{equation}
where ${{\bar {\cal D}}_n}=({\cal{R}}-{\cal {D}}_n)$ is the
complement of ${\cal {D}}_n$. Given that ${\cal {D}}$ is non-empty
(i.e., ${\cal {D}} ={\cal {D}}_n$), a RN R$_i$ is chosen from
${{\cal {D}}_n}$ to forward its decoded signal ${\hat x_i}$ to D.
Since all RNs within the decoding set ${\cal {D}}$ succeed in
perfectly decoding the SS $x$, we have ${\hat x_i} = x$
for $R_i \in {{\cal {D}}_n}$. Without loss of generality,
%considering that relay $ r_i  \in {\cal {D}}_n$ is
%selected to forward its decoded result $x$, we can express the
%received signal at destination as
%\begin{equation}\label{equa18}
%{y_d} = \sqrt P {h_{id}} x + {n_d}, \textrm{ } i \in {{\cal {D}}_n},
%\end{equation}
%where $h_{id}$ represents the fading coefficient of channel from
%relay $r_i$ to destination and $n_d$ represents zero-mean AWGN with
%variance $N_0$. Meanwhile, the received signal at eavesdropper is
%given by
%\begin{equation}\label{equa19} {y_e} = \sqrt P {h_{ie}} x +
%{n_e}, \textrm{ } i \in {{\cal {D}}_n},
%\end{equation}
%where $h_{ie}$ represents the fading coefficient of channel from
%relay $r_i$ to eavesdropper and $n_e$ represents zero-mean AWGN with
%variance $N_0$. Therefore,
given that $ {\cal {D}} = {\cal {D}}_n$ occurs and the RN R$_i \in
{{\cal {D}}_n}$ is selected, the corresponding R$_i$-D and R$_i$-E
channel capacities are
\begin{equation}\label{equa16} {C_{id}} = \frac{1}{2}{\log
_2}\left( {1 + |{h_{id}}{|^2}\gamma } \right),
\end{equation}
and
\begin{equation}\label{equa17}
{C_{ie}} = \frac{1}{2}{\log _2}\left( {1 + |{h_{ie}}{|^2}\gamma } \right),
\end{equation}
where $ R_i \in {{\cal {D}}_n}$. Naturally, it is wise
to rely on the specific RN having the highest channel capacity
${C_{id}}$, which is formulated as
\begin{equation}\label{equa18} {\textrm{Best Relay}} = \arg \mathop
{\max }\limits_{R_i \in {\cal {D}}_n } C_{id} = \arg \mathop {\max }\limits_{i
\in {\cal {D}}_n } |{h_{id}}{|^2}.
\end{equation}
Fortunately, since the WTC is typically independent of the main
channel, no WTC capacity gain is achieved by the ORS of (18), implying a beneficial PLS improvement.

By exploiting the ORS criterion of (18), a centralized or distributed relay selection (RS) algorithm may be developed [22]. To be specific, a centralized RS algorithm stores the {channel state information (CSI)} of the main channels, i.e, ${|h_{id}|^2 }$. Therefore, the best relay may be readily identified according to (18). By contrast, a distributed RS algorithm requires each RN to maintain a timer, whose initial value is set inverse-proportionally to ${|h_{id}|^2 }$ so that the best RN's timer is exhausted first. Once the best RN's timer is exhausted, it broadcasts a control packet to notify the other network nodes. Below, we present the SRT analysis of the proposed ORS scheme to quantify its advantages.

\subsection{SRT Analysis}
Again, for an empty decoding set $\cal {D}$, D is unable to decode
the SS. By contrast, if the decoding set is non-empty, a
RN will be chosen according to (18) for forwarding the SS to
D. Hence, using the law of total probability [22], we arrive at the
OP of the S-D link using the ORS scheme as
\begin{equation}\label{equa19}
P_{{\textrm{out}}}^{{\textrm{ORS}}} = \Pr ({\cal {D}} = \emptyset ) + \sum\limits_{n = 1}^{{2^N} - 1}{\Pr ({\cal {D}}
= {{\cal {D}}_n})\Pr (C_{bd}^{{\textrm{ORS}}} < R_d)},
\end{equation}
where ${C_{bd}^{{\textrm{ORS}}}}$ represents the channel capacity from
the best RN (denoted by R$_{\textrm{best}}$) to D. From (16) and (18),
${C_{bd}^{{\textrm{ORS}}}}$ may be expressed as
\begin{equation}\label{equa20}
C_{bd}^{{\textrm{ORS}}} = \mathop
{\max }\limits_{R_i \in {{\cal {D}}_n}} {C_{id}} = \frac{1}{2}{\log _2}\left(
{1 + \mathop {\max }\limits_{R_i \in {{\cal {D}}_n}} |{h_{id}}{|^2}{\gamma}}
\right).
\end{equation}
Since the $|h_{si}|^2$ factors of different RNs are independent of
each other and obey the exponential distribution with a mean of
$\sigma^2_{si}$, the probability of occurrence $\Pr ({\cal {D}} =
\emptyset )$ for ${\cal {D}} =\emptyset$ is obtained from (14) as
\begin{equation}\label{equa21}
\begin{split}
\Pr ({\cal {D}} = \emptyset ) &=\prod\limits_{i = 1}^N {\Pr \left( {\frac{1}{2}{{\log }_2}\left( {1
+ |{h_{si}}{|^2}{\gamma}} \right) < R_d} \right)}\\
&= \prod\limits_{i =1}^N {\Pr \left( {|{h_{si}}{|^2} < \delta } \right)}\\
&=\prod\limits_{i = 1}^N {\left[ {1 - \exp ( - \frac{\delta }{{\sigma
_{si}^2}})} \right]},
\end{split}
\end{equation}
where $N$ is the number of RNs and $\delta  = \frac{{{2^{2R_d}} -
1}}{{{\gamma}}}$.
%Notice that random variable $|h_{si}|^2$ is
%exponentially distributed with mean . Thus, the preceding equation
%can be further given by
%\begin{equation}\label{equa26} \Pr (D = \emptyset ) =
%\prod\limits_{i = 1}^N {\left[ {1 - \exp ( - \frac{\delta }{{\sigma
%_{si}^2}})} \right]}.
%\end{equation}
From (15), the probability of occurrence $\Pr ({\cal {D}} = {\cal
  {D}}_n )$ for the event ${\cal {D}} ={\cal {D}}_n$ is given by
\begin{equation}\label{equa22}
\begin{split}
\Pr ({\cal {D}} = {{\cal {D}}_n}) &= \prod\limits_{R_i \in {{\cal {D}}_n}} {\Pr \left( {\frac{1}{2}{{\log }_2}
\left( {1 + |{h_{si}}{|^2}{\gamma}} \right) > R_d} \right)}\\
&\quad\times \prod\limits_{R_j \in {{\bar {\cal D}}_n}}
{\Pr \left( {\frac{1}{2}{{\log }_2}\left( {1 + |{h_{sj}}{|^2}{\gamma}} \right) < R_d} \right)} \\
 &= \prod\limits_{R_i \in {{\cal {D}}_n}} {\Pr \left( {|{h_{si}}{|^2} > \delta } \right)} \prod\limits_{j
 \in {{\bar {\cal D}}_n}} {\Pr \left( {|{h_{sj}}{|^2} < \delta } \right)} \\
 &= \prod\limits_{R_i \in {{\cal {D}}_n}} {\exp ( - \frac{\delta }{{\sigma _{si}^2}})} \prod
 \limits_{R_j \in {{\bar {\cal D}}_n}} {\left[ {1 - \exp ( - \frac{\delta }{{\sigma _{sj}^2}})} \right]},
\end{split}
\end{equation}
where $\sigma _{si}^2 = E(|{h_{si}}{|^2})$ and $\sigma _{sj}^2 =
E(|{h_{sj}}{|^2})$. Additionally, we can obtain $\Pr
(C_{bd}^{{\textrm{ORS}}} < R_d)$ from (20) as
\begin{equation}\label{equa23}
\begin{split}
\Pr (C_{bd}^{{\textrm{ORS}}} < R_d) &= \Pr \left( {\mathop {\max }\limits_{R_i \in {{\cal {D}}_n}}
|{h_{id}}{|^2} < \delta } \right)\\
 &= \prod\limits_{R_i \in {{\cal {D}}_n}} {\left[ {1 - \exp
( - \frac{\delta }{{\sigma _{id}^2}})} \right]},
\end{split}
\end{equation}
where $\sigma _{id}^2 = E(|{h_{id}}{|^2})$. Hence, upon substituting
(21)-(23) into (19), we arrive at the closed-form OP expression of the ORS scheme. Meanwhile, E will also attempt to decode the SS based on its signals received from both S and the selected RN (if any). Recall that this is in contrast to the action of D, which only relies on the RN, since S is out of range for D. In this way, even when the
successful decoding set is empty and no RN forwards the SS, E might
still decode the SS. Moreover, if the successful decoding set is
non-empty, E will overhear the transmissions of both S as well as of
the selected RN and performs detection using both received signal
copies. By using the selection diversity combining, the capacity
achieved by E for ${\cal {D}} ={\cal {D}}_n$ with the aid of the ORS
scheme is the higher one of $C_{se}$ and $C_{be}$, yielding
\begin{equation}\label{equa24}
C_e^{{\textrm{ORS}}} = \max \left( {C_{se}^{{\textrm{ORS}}},C_{be}^{{\textrm{ORS}}}} \right),
\end{equation}
where ${C^{\textrm{ORS}}_{se}}$ and $C_{be}^{{\textrm{ORS}}}$,
respectively, represent the S-E and R$_{\textrm{best}}$-E capacities
given by
\begin{equation}\label{equa25}
C_{se}^{{\textrm{ORS}}} = \frac{1}{2}{\log _2}\left( {1 + |{h_{se}}{|^2}{\gamma}} \right),
\end{equation}
and
\begin{equation}\label{equa26}
C_{be}^{{\textrm{ORS}}} = \frac{1}{2}{\log _2}\left( {1 + |{h_{be}}{|^2}{\gamma}} \right),
\end{equation}
where $|h_{be}|^2$ represents the R$_{\textrm{best}}$-E fading
coefficient. Hence, using the law of total probability, we obtain
the IP at E as
\begin{equation}\label{equa27}
\begin{split}
P_{{\textrm{int}}}^{{\textrm{ORS}}} = &\Pr ({\cal {D}} = \emptyset )\Pr
(C_{se}^{{\textrm{ORS}}} > R_d) \\
&+ \sum\limits_{n = 1}^{{2^N} - 1} {\Pr
({\cal {D}} = {{\cal {D}}_n})\Pr (C_e^{{\textrm{ORS}}} > R_d)},
\end{split}
\end{equation}
where $\Pr ({\cal {D}} = \emptyset )$ and $\Pr ({\cal {D}} = {\cal
{D}}_n )$ are given by (21) and (22), respectively. Using (25), the
term $\Pr (C_{se}^{{\textrm{ORS}}} > R_d)$ is readily obtained as
\begin{equation}\label{equa28}
\Pr (C_{se}^{{\textrm{ORS}}} > R_d) = \Pr \left( {|{h_{se}}{|^2} > \delta } \right)
= \exp ( - \frac{\delta }{{\sigma _{se}^2}}).
\end{equation}
Additionally, for ${\cal {D}} ={\cal {D}}_n$, we obtain $\Pr
(C_e^{{\textrm{ORS}}}
> R_d)$ from Appendix A as (29) at the top of the following page,
\begin{figure*}
\begin{equation}\label{equa29}
\begin{split}
\Pr (C_e^{{\textrm{ORS}}} > R_d) = \sum\limits_{R_i \in {{\cal {D}}_n}} {\left[ {1 +
\sum\limits_{k = 1}^{{2^{|{{\cal {D}}_n}| - 1}} - 1} {{{( - 1)}^{|{{\cal{A}}_k}|}}{(1 +
\sum\limits_{R_j \in {A_k}} {\frac{{\sigma _{id}^2}}{{\sigma _{jd}^2}}} )^{ - 1}}} } \right]}\left[ {\exp ( - \frac{\delta }{{\sigma _{ie}^2}}) +
\exp ( - \frac{\delta }{{\sigma _{se}^2}}) - \exp ( - \frac{\delta }{{\sigma _{ie}^2}}
- \frac{\delta }{{\sigma _{se}^2}})} \right]
\end{split}
\end{equation}
\end{figure*}
where $|{{\cal {D}}_n}|$ represents the cardinality of the set ${\cal  {D}}_n$, ${{\cal{A}}_k}$ is the $k$-th non-empty subset of $\{{{\cal  {D}}_n-i}\}$, and $|{{\cal{A}}_k}|$ is the cardinality of the set  ${{\cal{A}}_k}$. Thus, a closed-form IP expression of the ORS scheme is derived by substituting (21), (22), (28) and (29) into (27). So far, both the OP and IP have been derived in (19) and (27), which characterize the SRT for the ORS scheme. In order to further simplify the OP and IP expressions, we now consider a special case, where the fading coefficients of all main links (i.e., $|h_{sd}|^2$, $|h_{si}|^2$, and $|h_{id}|^2$) are independent and identically distributed (i.i.d.) random variables having the same average
channel gain of $\sigma^2_m$. This assumption is valid in a statistical sense when all RNs are mobile and uniformly distributed around S and D. Moreover, if the main links have different average channel gains, we can use Eqs. (19) and (27) to quantify the SRT of proposed ORS scheme. Similarly, the fading coefficients of all WTCs (i.e., $|h_{se}|^2$ and $|h_{ie}|^2$) are also assumed to be i.i.d. random variables having the same average channel gain of $\sigma^2_e$. Let $\lambda_{me}=\sigma^2_{m}/\sigma^2_{e}$ denote the ratio of $\sigma^2_{m}$ to $\sigma^2_{e}$, which we refer to as the main-to-eavesdropper ratio (MER) throughout this paper. Hence, considering $\sigma^2_{si}=\sigma^2_m$, we can simplify (21) and (22) to
\begin{equation}\label{equa30}
\Pr ({\cal {D}} = \emptyset ) = {\left[ {1 - \exp ( - \frac{\delta }{{\sigma _{m}^2}})} \right]^N}.
\end{equation}
and
\begin{equation}\label{equa31}
\Pr ({\cal {D}} = {{\cal {D}}_n}) = \exp ( - \frac{{|{{\cal {D}}_n}|\delta }}{{\sigma _{m}^2}}){\left[ {1 -
\exp ( - \frac{\delta }{{\sigma _{m}^2}})} \right]^{|{{\bar {\cal D}}_n}|}},
\end{equation}
where $|{\cal {D}}_n|$ and $|\bar {\cal {D}}_n|$ are the
cardinalities of the sets ${\cal {D}}_n$ and $\bar {\cal {D}}_n$,
respectively. Then, upon using $\sigma^2_{id}=\sigma^2_m$, we
rewrite (23) as
\begin{equation}\label{equa32} \Pr (C_{bd}^{{\textrm{ORS}}} < R_d) =
{\left[ {1 - \exp ( - \frac{\delta }{{\sigma _m^2}})} \right]^{|{{
D}_n}|}}.
\end{equation}
Substituting (30)-(32) into (19) yields
\begin{equation}\label{equa33}
\begin{split}
P_{{\textrm{out}}}^{{\textrm{ORS}}}
%& = {\left[ {1 - \exp ( - \frac{\delta }{{\sigma _m^2}})}
%\right]^N} + \sum\limits_{n = 1}^{{2^N} - 1} {\exp ( - \frac{{|{{\cal {D}}_n}|\delta }}{{\sigma _m^2}})
%{{\left[ {1 - \exp ( - \frac{\delta }{{\sigma _m^2}})} \right]}^N}} \\
&= {\left[ {1 - \exp ( - \frac{\delta }{{\sigma _m^2}})} \right]^N}\left[ {1 + \sum\limits_{n = 1}
^{{2^N} - 1} {\exp ( - \frac{{|{{\cal {D}}_n}|\delta }}{{\sigma _m^2}})} } \right]\\
&= {\left[ {1 - \exp ( - \frac{\delta }{{\sigma _m^2}})} \right]^N}{\left[ {1 + \exp ( - \frac{\delta }
{{\sigma _m^2}})} \right]^N}\\
&= {\left[ {1 - \exp ( - \frac{{2\delta }}{{\sigma _m^2}})} \right]^N}.
\end{split}
\end{equation}
Meanwhile, upon considering
$\sigma^2_{se}=\sigma^2_{ie}=\sigma^2_e$, we can rewrite (28) and
(29) as
\begin{equation}\label{equa34}
\Pr (C_{se}^{{\textrm{ORS}}} > R_d) =\exp ( - \frac{\delta }{{\sigma _e^2}}),
\end{equation}
and
\begin{equation}\label{equa35}
\Pr (C_e^{{\textrm{ORS}}} > R_d) =2\exp ( - \frac{\delta }{{\sigma _e^2}})
- \exp ( - \frac{{2\delta }}{{\sigma _e^2}}).
\end{equation}
Substituting (30), (31), (34) and (35) into (27), we arrive at the
IP as
\begin{equation}\label{equa36}
\begin{split}
P_{{\textrm{int}}}^{{\textrm{ORS}}} =& {\left[ {1 - \exp ( - \frac{\delta }{{\sigma _m^2}})} \right]^N}
\exp ( - \frac{\delta }{{\sigma _e^2}})\\
 &+ \left[ {1 - {{\left( {1 - \exp ( - \frac{\delta }{{\sigma _m^2}})} \right)}^N}} \right]\\
&\quad \times\left[ {2
 \exp ( - \frac{\delta }{{\sigma _e^2}}) - \exp ( - \frac{{2\delta }}{{\sigma _e^2}})} \right].
\end{split}
\end{equation}
Combining (33) and (36) yields
\begin{equation}\label{equa37}
\begin{split}
P_{{\textrm{int}}}^{{\textrm{ORS}}} =& {\left( {1 - {\theta ^{1/2}}} \right)^N}
{\theta ^{{\lambda _{me}}/2}} \\
&+ \left[ {1 - {{\left( {1 - {\theta ^{1/2}}} \right)}^N}}
\right]\left( {2{\theta ^{{\lambda _{me}}/2}} - {\theta ^{{\lambda _{me}}}}} \right),
\end{split}
\end{equation}
where $\theta  = 1 - {(P_{{\textrm{out}}}^{{\textrm{ORS}}})^{1/N}}$.
Let us now analyze the SRT of our ORS scheme, as the number of RNs
$N$ tends to infinity. Observe from (30) that as $N \to \infty $,
the probability of occurrence for the event ${\cal {D}} =\emptyset$
tends to zero, i.e., we have $\Pr ({\cal {D}} = \emptyset ) = 0$ for
$N \to \infty $. Substituting this result into (36) gives
\begin{equation}\label{equa38}
P_{{\textrm{int}}}^{{\textrm{ORS}}} = 2\exp ( - \frac{\delta }{{\sigma _e^2}}) - \exp
( - \frac{{2\delta }}{{\sigma _e^2}}),
\end{equation}
for $N \to \infty $. Thus, letting $N \to \infty $ and combining
(33) and (38), we arrive at
\begin{equation}\label{equa39}
P_{{\textrm{int}}}^{{\textrm{ORS}}} = 2{\left[ {1 - {{(P_{{\textrm{out}}}^{{\textrm{ORS}}})}^{1/N}}}
\right]^{{\lambda _{me}}/2}} - {\left[ {1 - {{(P_{{\textrm{out}}}^{{\textrm{ORS}}})}^{1/N}}} \right]
^{{\lambda _{me}}}}.
\end{equation}
Observe from (39) that given a specific OP constraint $0 <
P_{{\textrm{out}}}^{{\textrm{ORS}}} < 1$, the IP
$P_{{\textrm{int}}}^{{\textrm{ORS}}}$ of our ORS scheme
asymptotically tends to zero for $N \to \infty$. Additionally, using
(33) and (38), we can rewrite $P_{{\textrm{out}}}^{{\textrm{ORS}}}$
as a function of $P_{{\textrm{int}}}^{{\textrm{ORS}}}$, yielding
\begin{equation}\label{equa40}
P_{{\textrm{out}}}^{{\textrm{ORS}}} = {\left[ {1 - {{\left( {1 - \sqrt {1 - P_{{\textrm{int}}}^
{{\textrm{ORS}}}} } \right)}^{2\lambda _{me}^{ - 1}}}} \right]^N},
\end{equation}
which shows that for an IP constraint of $0 <
P_{{\textrm{int}}}^{{\textrm{ORS}}} < 1$, the OP
$P_{{\textrm{out}}}^{{\textrm{ORS}}}$ of our ORS scheme also tends
to zero for $N \to \infty$. In other words, given the maximal
tolerable IP, the ORS scheme minimizes the OP as $N \to \infty$.
Conversely, as seen in (39), given the maximal tolerable OP, the ORS
scheme minimizes the IP for $N \to \infty$. Additionally, the SRT of
the ORS scheme depends not only on the number of RNs $N$ but also on
the average channel gains $\sigma^2_{si}$, $\sigma^2_{id}$,
$\sigma^2_{se}$ and $\sigma^2_{ie}$ of the main links and the WTCs.
Given the maximal tolerable IP and OP, we can directly determine the
number of RNs required, provided that the average channel gains of
$\sigma^2_{si}$, $\sigma^2_{id}$, $\sigma^2_{se}$ and
$\sigma^2_{ie}$ are known. It has to be pointed out that the average
channel gain is typically dominated by the path loss exponent and
the transmission distance. Assuming that the eavesdroppers are
uniformly distributed around S within the coverage area, we can
determine the average S-E distance and then use the distance to
estimate the average channel gain $\sigma^2_{se}$ for a specific
path-loss model. Furthermore, the average R-E channel gain
$\sigma^2_{ie}$ can be similarly estimated. Moreover, the average
channel gains $\sigma^2_{si}$ and $\sigma^2_{id}$ of the main links
may be determined by averaging out the fading coefficients
$|h_{si}|^2$ and $|h_{id}|^2$. Once the average channel gains of
$\sigma^2_{si}$, $\sigma^2_{id}$, $\sigma^2_{se}$ and
$\sigma^2_{ie}$ have been obtained, we can determine the number of
RNs required for maintaining the desired SRT.

\section{Numerical Results and Discussions}
\begin{figure}
  \centering
  {\includegraphics[scale=0.55]{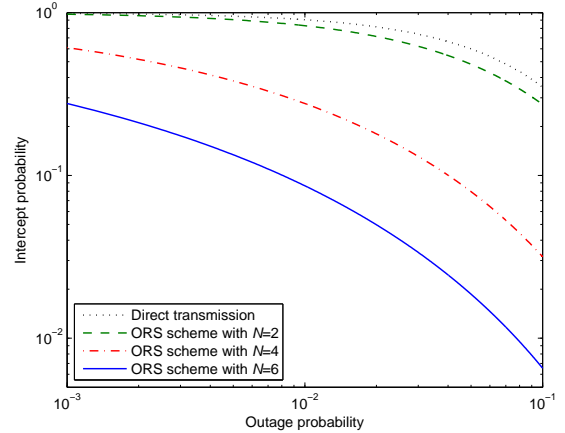}\\
  \caption{{SRT of the traditional DT and of the ORS schemes
  for different number of RNs associated with an MER of $\lambda_{me}=10\textrm{dB}$.}}\label{Fig3}}
\end{figure}

\begin{figure}
  \centering
  {\includegraphics[scale=0.55]{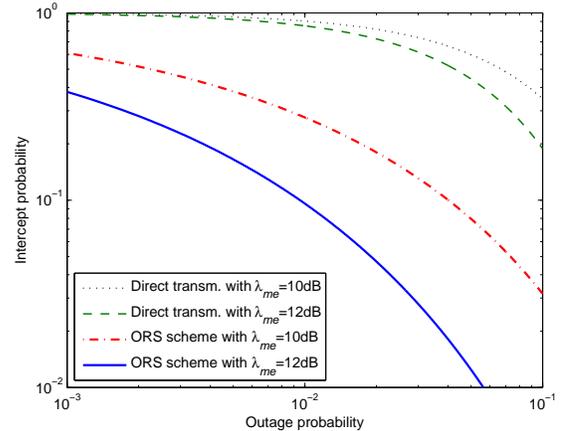}\\
  \caption{{SRT of the traditional DT and of the ORS schemes
  for different MERs $\lambda_{me}$ with $N=4$.}}\label{Fig4}}
\end{figure}

In Fig. 3, we show our numerical SRT results for both the conventional direct transmission (DT) and the proposed ORS schemes for different number of RNs at $\lambda_{me}=10\textrm{dB}$. As shown in Fig. 3, when the OP degrades from $10^{-3}$ to $10^{-1}$, the IP of both schemes improves. One can also see from Fig. 3 that for a specific OP, the IP of the ORS scheme corresponding to $N=2$, $N=4$ and $N=6$ is strictly lower than that of DT. This confirms that our ORS scheme performs better than the conventional DT in terms of SRT. In contrast to Fig. 3, Fig. 4 shows the IP versus OP of both the traditional DT and the ORS schemes for different MERs $\lambda_{me}$ associated with $N=4$. Observe from Fig. 4 that for both $\lambda_{me}=10{\textrm{dB}}$ and $\lambda_{me}=12{\textrm{dB}}$, the ORS scheme strictly outperforms the DT.

Fig. 5 shows the OP versus the number of RNs $N$ of the ORS scheme for different IP constraints associated with MER $\lambda_{me}=5\textrm{dB}$. As shown in Fig. 5, when the IP increases from $P_{\textrm{int}}=10^{-3}$ to $P_{\textrm{int}}=10^{-1}$, the OP of the ORS scheme is significantly reduced. Additionally, we can also observe from Fig. 5 that for $P_{\textrm{int}}=10^{-1}$, $P_{\textrm{int}}=10^{-2}$ and $P_{\textrm{int}}=10^{-3}$, the OP of the ORS scheme tends to zero, as the number of RNs increases from $N=1$ to $N=10^3$, demonstrating the reliability improvement upon increasing the number of RNs.

\begin{figure}
  \centering
  {\includegraphics[scale=0.55]{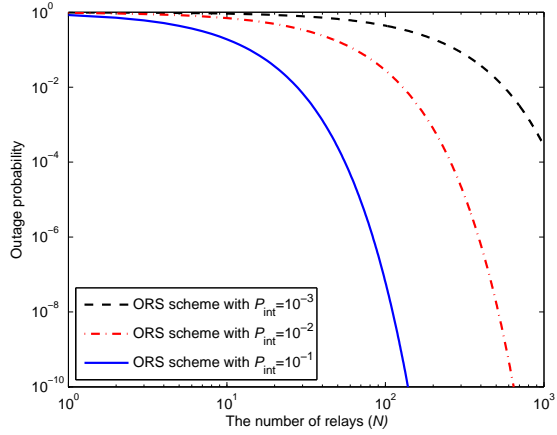}\\
  \caption{OP versus the number of RNs $N$ of the ORS scheme
  for different IP constraints associated with MER $\lambda_{me}=5\textrm{dB}$.}\label{Fig5}}
\end{figure}

Fig. 6 shows the IP versus the number of RNs $N$ for the ORS scheme
under different OP constraints at an MER of $\lambda_{me}=5\textrm{dB}$. Observe from Fig. 6 that as the OP increases from $P_{\textrm{out}}=10^{-3}$ to $P_{\textrm{out}}=10^{-1}$, the IP is reduced, which further confirms that the grade of PLS improves, as the OP requirements are relaxed. Fig. 6 also shows that for all the cases of $P_{\textrm{out}}=10^{-1}$, $P_{\textrm{out}}=10^{-2}$ and
$P_{\textrm{out}}=10^{-3}$, the IP of the ORS scheme decreases as
the number of RNs increases from $N=1$ to $N=10^4$. This implies
that given a maximal tolerable OP, the IP of the ORS scheme tends to
zero, as $N \to \infty$.

\begin{figure}
  \centering
  {\includegraphics[scale=0.55]{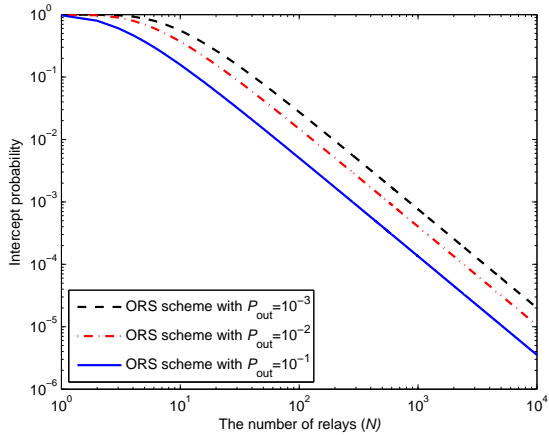}\\
  \caption{IP versus the number of RNs $N$ of the ORS scheme
  for different OP constraints associated with MER $\lambda_{me}=5\textrm{dB}$.}\label{Fig6}}
\end{figure}

\section{Conclusions}
We have investigated the SRT of wireless communications in the
presence of eavesdropping attacks. We have derived the
SRT of conventional DT in the presence of an eavesdropper over
Rayleigh fading channels and shown that the IP may be improved by
relaxing the OP requirement and vice versa. Additionally, we have
quantified the benefits of ORS relying on DF relaying in Rayleigh
fading environments. Our numerical results have shown that the ORS
scheme strictly outperforms the conventional DT scheme in terms of
its SRT. Finally, as the number of RNs increases, the SRT
performance of the ORS scheme significantly improves, demonstrating
the security and reliability benefits of relying on multiple RNs.

{Here we only studied the single-source, single-destination and single-eavesdropper scenario relying on the assistance of multiple relays in wireless networks. In our future research we will consider the extension of this work to a general scenario of multi-source, multi-destination, and multi-eavesdropper situations, where cooperative beamforming may be adopted to protect the legitimate transmission against multiple eavesdroppers. To be specific, multiple source-destination pairs can collaborate with each other to form virtual antenna arrays for optimum beamforming so that the desired signals received at legitimate receivers experience constructive interference. By contrast, the illegitimate eavesdroppers would be subjected to destructive interference. Therefore, with the cooperative beamforming, the received signal strength of legitimate receivers would be much higher than that of the eavesdroppers, leading to a significant wireless security enhancement.

\appendices
\section{Derivation of (29)}
Using (24), we obtain the term $\Pr (C_e^{{\textrm{ORS}}} > R_d)$ as
\begin{equation}\nonumber
\Pr (C_e^{{\textrm{ORS}}} > R_d) = \Pr \left[ {\max \left( {C_{se}^{{\textrm{ORS}}},C_{be}
^{{\textrm{ORS}}}} \right) > R_d} \right].
\label{A.1}\tag{A.1}
\end{equation}
Substituting $C_{se}^{{\textrm{ORS}}}$ and $C_{be}^{{\textrm{ORS}}}$
from (25) and (26) into (A.1) yields
\begin{equation}\nonumber
\Pr (C_e^{{\textrm{ORS}}} > R_d) = 1 - \Pr \left( {|{h_{se}}{|^2} < \delta } \right)\Pr
\left( {|{h_{be}}{|^2} < \delta } \right),\label{A.2}\tag{A.2}
\end{equation}
where $\delta  = \frac{{{2^{2R_d}} - 1}}{{{\gamma}}}$ and
$|h_{se}|^2$ and $|h_{be}|^2$ represent the fading coefficients of
the S-E and R$_{\textrm{best}}$-E, respectively. Since $|h_{se}|^2$
is an exponentially distributed random variable with a mean of
$\sigma^2_{se}$, we have
\begin{equation}\nonumber
\Pr \left( {|{h_{se}}{|^2} < \delta } \right) = 1 - \exp ( -
\frac{\delta }{{\sigma _{se}^2}}).\label{A.3}\tag{A.3}
\end{equation}
Again in the ORS scheme, the best relay is determined according to
(18) and any of the RNs within the decoding set ${\cal {D}}_n$ may
become the best relay, provided that we have $|h_{id}|^2>|h_{jd}|^2$
for $R_j \in \{{\cal {D}}_n-i\}$, in which `$-$' represents the
difference set. Thus, using the law of total probability, the term
$\Pr \left( {|{h_{be}}{|^2} < \delta } \right) $ may be expressed as
\begin{equation}\nonumber
\begin{split}
&\Pr \left( {|{h_{be}}{|^2} < \delta } \right) \\
&= \sum\limits_{R_i \in {{\cal {D}}_n}} \Pr \left( {|{h_{id}}{|^2}
 > \mathop {\max }\limits_{R_j \in \{ {{\cal {D}}_n} - i\} } |{h_{jd}}{|^2}} \right)\\
&\quad \quad \quad \quad \times \Pr \left( {|{h_{ie}}{|^2} <
  \delta} \right).
  \end{split}\label{A.4}\tag{A.4}
\end{equation}
Upon assuming $|h_{id}|^2=x$, we can rewrite ${\Pr \left(
{|{h_{id}}{|^2}
> \mathop {\max }\limits_{R_j \in \{ {{\cal {D}}_n} - i\} } |{h_{jd}}{|^2}}
\right)}$ as $\Pr \left( {\mathop {\max }\limits_{R_j \in \{ {{\cal
{D}}_n} - i\} } |{h_{jd}}{|^2} < x} \right) $, which is further
reformulated as (A.5) at the top of the following page,
\begin{figure*}
\begin{equation}\nonumber
\begin{split}
\Pr \left( {\mathop {\max }\limits_{R_j \in \{ {{\cal {D}}_n} - i\} } |{h_{jd}}{|^2} < x} \right) &= \int_0^\infty
 {\prod\limits_{R_j \in \{ {{\cal {D}}_n} - i\} } {[1 - \exp ( - \frac{x}{{\sigma _{jd}^2}})]} \frac{1}{{\sigma _{id}^2}}
 \exp ( - \frac{x}{{\sigma _{id}^2}})dx} \\
&= \int_0^\infty  {[1 + \sum\limits_{k = 1}^{{2^{|{{\cal {D}}_n}| - 1}} - 1} {{{( - 1)}^{|{{\cal{A}}_k}|}}\exp ( -
\sum\limits_{R_j \in {{\cal{A}}_k}} {\frac{x}{{\sigma _{jd}^2}}} )} ]\frac{1}{{\sigma _{id}^2}}\exp ( - \frac{x}
{{\sigma _{id}^2}})dx} \\
&= 1 + \sum\limits_{k = 1}^{{2^{|{{\cal {D}}_n}| - 1}} - 1} {{{( - 1)}^{|{{\cal{A}}_k}|}}\int_0^\infty  {\frac{1}
{{\sigma _{id}^2}}\exp ( - \frac{x}{{\sigma _{id}^2}} - \sum\limits_{R_j \in {{\cal{A}}_k}} {\frac{x}
{{\sigma _{jd}^2}}} )dx} } \\
&= 1 + \sum\limits_{k = 1}^{{2^{|{{\cal {D}}_n}| - 1}} - 1} {{{( - 1)}^{|{{\cal{A}}_k}|}}{(1 +
\sum\limits_{R_j \in {{\cal{A}}_k}} {\frac{{\sigma _{id}^2}}{{\sigma _{jd}^2}}} )^{ - 1}}},
\end{split}\label{A.5}\tag{A.5}
\end{equation}
\end{figure*}
where $|{{\cal {D}}_n}|$ represents the cardinality of the set
${\cal {D}}_n$, ${{\cal{A}}_k}$ is the $k$-th non-empty subset of
$\{{{\cal {D}}_n-i}\}$ and $|{{\cal{A}}_k}|$ represents the
cardinality of the set ${{\cal{A}}_k}$. Additionally, considering
that $|h_{ie}|^2$ is an exponentially distributed random variable,
we have
\begin{equation}\nonumber \Pr \left( {|{h_{ie}}{|^2} < \delta
} \right) = 1 - \exp ( - \frac{\delta }{{\sigma
_{ie}^2}}),\label{A.6}\tag{A.6}
\end{equation}
where $\sigma _{ie}^2 = E(|{h_{ie}}{|^2})$. Substituting (A.5) and (A.6) into (A.4) gives (A.7) at the top of the following page.
\begin{figure*}
\begin{equation}\nonumber
\Pr \left( {|{h_{be}}{|^2} < \delta } \right) = \sum\limits_{R_i \in {{\cal {D}}_n}} {\left[ {1
+ \sum\limits_{k = 1}^{{2^{|{{\cal {D}}_n}| - 1}} - 1} {{{( - 1)}^{|{{\cal{A}}_k}|}}{(1 + \sum\limits_{j
\in {{\cal{A}}_k}} {\frac{{\sigma _{id}^2}}{{\sigma _{jd}^2}}} )^{ - 1}}} } \right]\left[ {1 - \exp
 ( - \frac{\delta }{{\sigma _{ie}^2}})} \right]}.\label{A.7}\tag{A.7}
\end{equation}
\end{figure*}
Thus, substituting $\Pr \left( {|{h_{se}}{|^2} < \delta } \right)$ and $\Pr \left( {|{h_{be}}{|^2}
 < \delta } \right)$ from (A.3) and (A.7) into (A.2), we easily obtain (29).
%\begin{equation}\nonumber
%\begin{split}
%\Pr (C_e^{{\textrm{ORS}}} > R_d) &= 1 - \sum\limits_{R_i \in {{\cal {D}}_n}} {\left[ {1 + \sum\limits_{k = 1}
%^{{2^{|{{\cal {D}}_n}| - 1}} - 1} {{{( - 1)}^{|{{\cal{A}}_k}|}}{(1 + \sum\limits_{R_j \in {A_k}} {\frac{{\sigma _{id}^2}}
%{{\sigma _{jd}^2}}} )^{ - 1}}} } \right]}\\
%&\quad\quad\quad\quad \times\left[ {1 - \exp ( - \frac{\delta }{{\sigma _{ie}^2}})} \right]\left[ {1 -
%\exp ( - \frac{\delta }{{\sigma _{se}^2}})} \right]\\
%&= \sum\limits_{R_i \in {{\cal {D}}_n}} {\left[ {1 + \sum\limits_{k = 1}^{{2^{|{{\cal {D}}_n}| - 1}} - 1} {{{( - 1)}^{|
%{{\cal{A}}_k}|}}{(1 + \sum\limits_{R_j \in {A_k}} {\frac{{\sigma _{id}^2}}{{\sigma _{jd}^2}}} )^{ - 1}}} }
%\right]} \\
%&\quad\quad \quad\times \left[ {\exp ( - \frac{\delta }{{\sigma _{ie}^2}}) + \exp ( - \frac{\delta }
%{{\sigma _{se}^2}}) - \exp ( - \frac{\delta }{{\sigma _{ie}^2}} - \frac{\delta }{{\sigma _{se}^2}})} \right],
%\end{split}\label{A.8}\tag{A.8}
%\end{equation}
%which is (29).

\begin{IEEEbiography}[{\includegraphics[width=1in,height=1.25in]{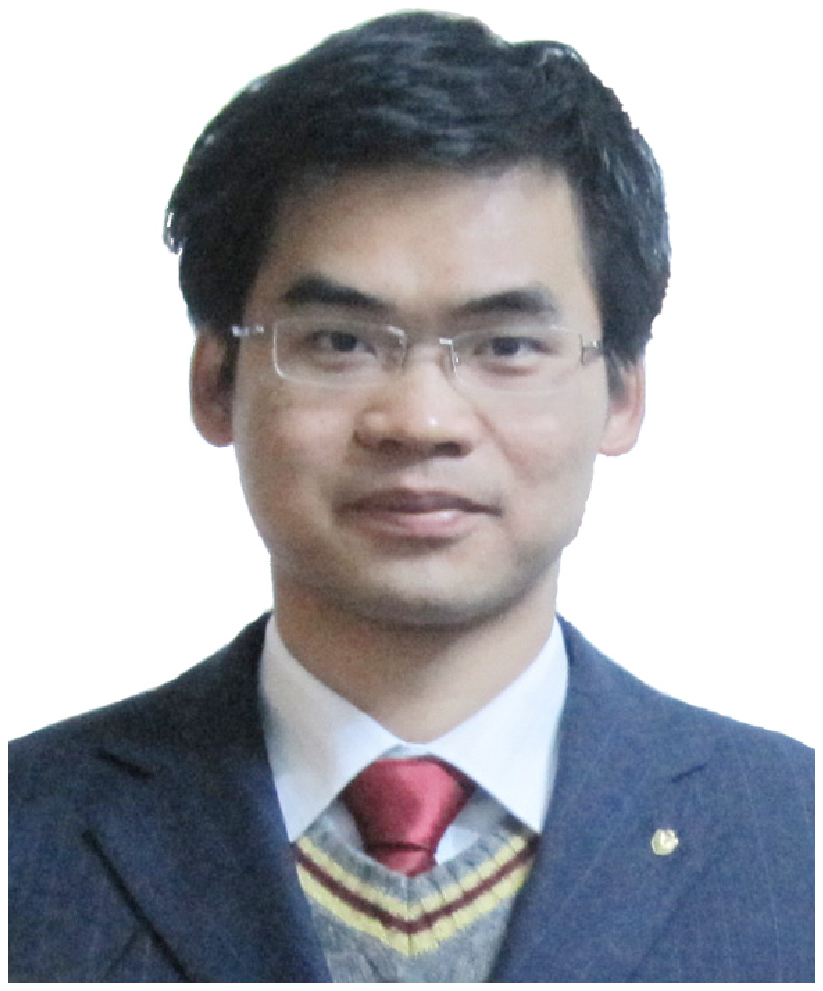}}]{Yulong
Zou} (S'07-M'12-SM'13) received the B.Eng. degree in information
engineering from the Nanjing University of Posts and
Telecommunications (NUPT), Nanjing, China, in July 2006, the first Ph.D. degree from the Stevens Institute of Technology, New Jersey, United States, in May 2012, and the second Ph.D. degree from NUPT, Nanjing, China, in July 2012.

Dr. Zou is currently serving as an editor for the IEEE
Communications Surveys \& Tutorials, IEEE Communications Letters,
EURASIP Journal on Advances in Signal Processing, and KSII
Transactions on Internet and Information Systems. He is also serving
as the lead guest editor for a special issue on ``Security Challenges and Issues in Cognitive Radio Networks" in the EURASIP Journal on
Advances in Signal Processing. In addition, he has acted as
symposium chairs, session chairs, and TPC members for a number of
IEEE sponsored conferences including the IEEE Wireless
Communications and Networking Conference (WCNC), IEEE Global
Telecommunications Conference (GLOBECOM), IEEE International
Conference on Communications (ICC), IEEE Vehicular Technology
Conference (VTC), International Conference on Communications in
China (ICCC), and so on.

His research interests span a wide range of topics in wireless
communications and signal processing, including the cooperative
communications, cognitive radio, wireless security, and green
communications. In these areas, he has published extensively in
internationally renowned journals, including the IEEE Transactions on Signal Processing, IEEE Transactions on Communications, IEEE Journal on Selected Areas in Communications, IEEE Transactions on Wireless Communications, IEEE Transactions on Vehicular Technology, and IEEE Communications Magazine.

\end{IEEEbiography}

\begin{IEEEbiography}[{\includegraphics[width=1in,height=1.25in]{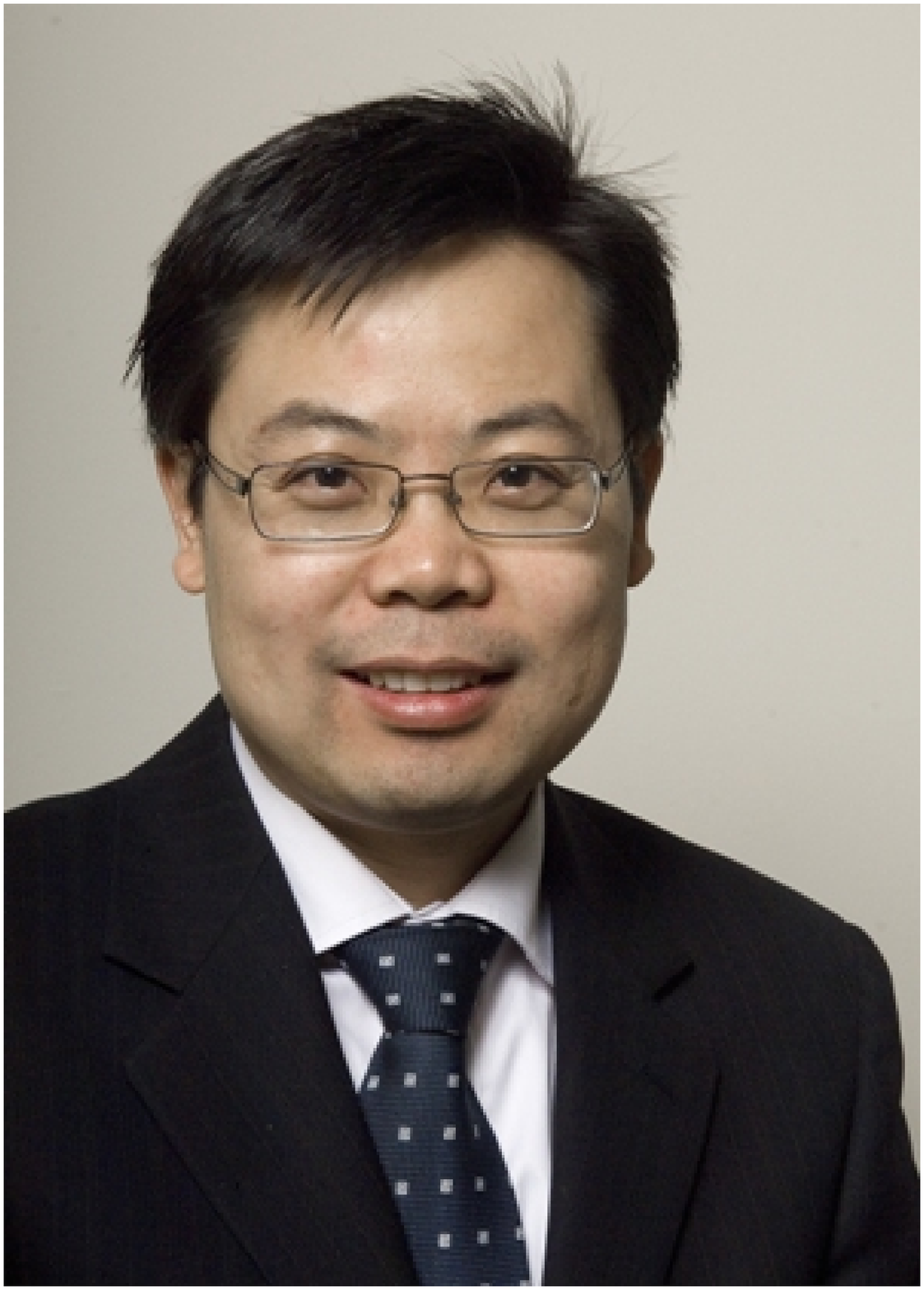}}]
{Xianbin Wang}(S'98-M'99-SM'06) is an Associate Professor at The
University of Western Ontario and a Canada Research Chair in
Wireless Communications. He received his Ph.D. degree in electrical
and computer engineering from National University of Singapore in
2001.

Prior to joining Western, he was with Communications Research Centre
Canada as Research Scientist/Senior Research Scientist between July
2002 and Dec. 2007. From Jan. 2001 to July 2002, he was a system
designer at STMicroelectronics, where he was responsible for system
design for DSL and Gigabit Ethernet chipsets. He was with Institute
for Infocomm Research, Singapore (formerly known as Centre for
Wireless Communications), as a Senior R \& D engineer in 2000. His
primary research area is wireless communications and related
applications, including adaptive communications, wireless security,
and wireless infrastructure based position location. Dr. Wang has
over 150 peer-reviewed journal and conference papers on various
communication system design issues, in addition to 23 granted and
pending patents and several standard contributions.

Dr. Wang is an IEEE Distinguished Lecturer and a Senior Member of
IEEE. He was the recipient of three IEEE Best Paper Awards. He
currently serves as an Associate Editor for IEEE Wireless
Communications Letters, IEEE Transactions on Vehicular Technology
and IEEE Transactions on Broadcasting. He was also an editor for
IEEE Transactions on Wireless Communications between 2007 and 2011.
Dr. Wang was involved in a number of IEEE conferences including
GLOBECOM, ICC, WCNC, VTC, and ICME, on different roles such as
symposium chair, track chair, TPC and session chair.

\end{IEEEbiography}

\begin{IEEEbiography}[{\includegraphics[width=1in,height=1.25in]{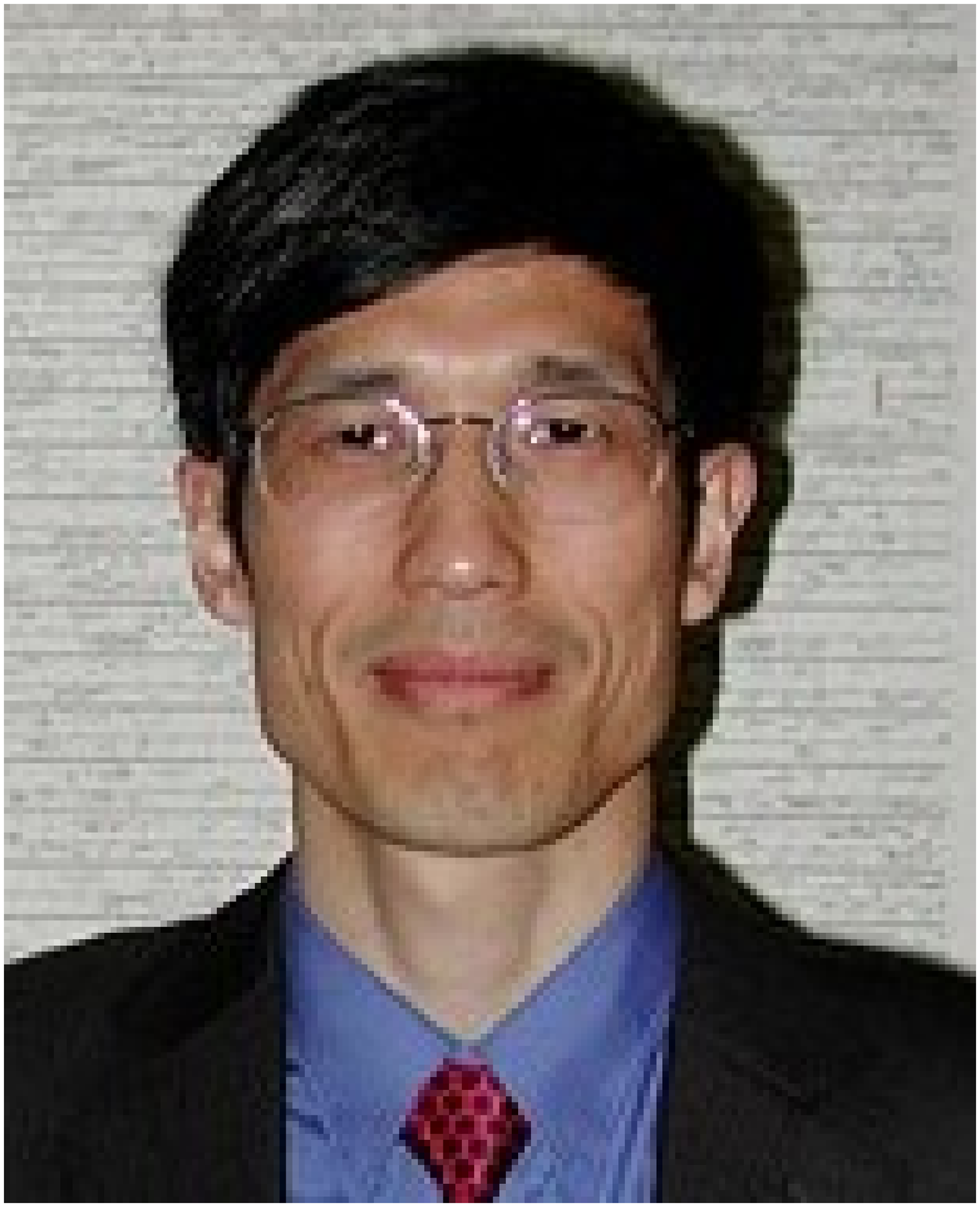}}]{Weiming Shen} is a Senior Research Scientist at the National Research Council Canada and an Adjunct Professor at Tongji University, China, and University of Western Ontario, Canada. He is a Fellow of IEEE. He received his Bachelor and Master¡¯s degrees from Northern (Beijing) Jiaotong University, China and his PhD degree from the University of Technology of Compi¨¨gne, France. His recent research interest includes agent-based collaboration technology and applications, wireless sensor networks. He has published several books and over 300 papers in scientific journals and international conferences in the related areas. His work has been cited over 6,000 times with an h-index of 37. He has been invited to provide over 60 invited lectures/seminars at different academic and research institutions over the world and keynote presentations / tutorials at various international conferences. He is a member of the Steering Committee for the IEEE Transactions on Affective Computing and an Associate Editor or Editorial Board Member of ten international journals (including IEEE Transactions on Automation Science and Engineering, Computers in Industry; Advanced Engineering Informatics; Service Oriented Computing and Applications) and served as guest editor for several other international journals.
\end{IEEEbiography}

\begin{IEEEbiography}[{\includegraphics[width=1in,height=1.25in]{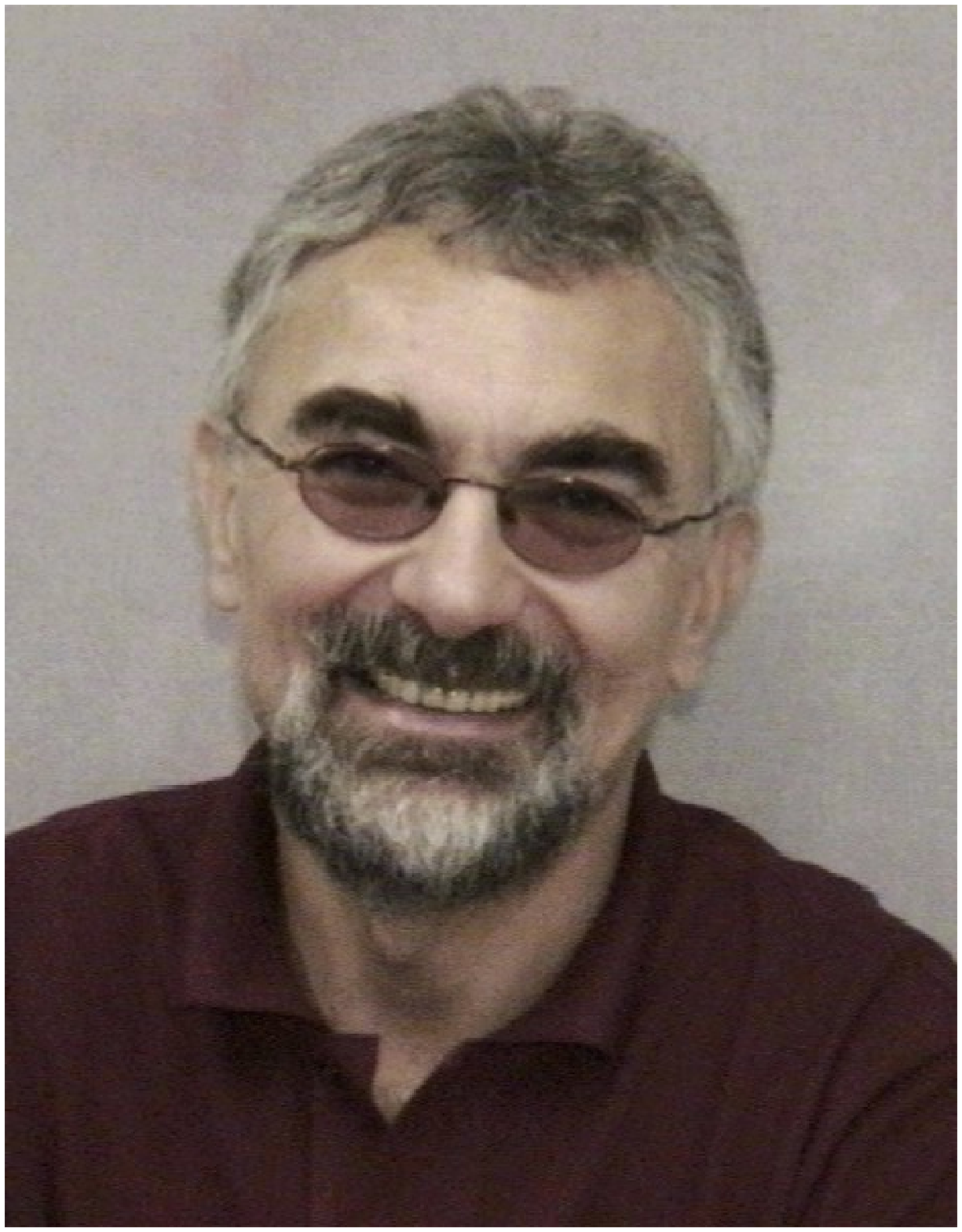}}]{Lajos Hanzo} (http://www-mobile.ecs.soton.ac.uk) FREng, FIEEE, FIET, Fellow of EURASIP, DSc received his degree in electronics in 1976 and his doctorate in 1983.  In 2009 he was awarded the honorary doctorate ``Doctor Honoris Causa'' by the Technical University of Budapest.  During his 37-year career in telecommunications he has held various research and academic posts in Hungary, Germany and the UK. Since 1986 he has been with the School of Electronics and Computer Science, University of Southampton, UK, where he holds the chair in telecommunications.  He has successfully supervised 80+ PhD students, co-authored 20 John Wiley/IEEE Press books on mobile radio communications totalling in excess of 10 000 pages, published 1300+ research entries at IEEE Xplore, acted both as TPC and General Chair of IEEE conferences, presented keynote lectures and has been awarded a number of distinctions. Currently he is directing a 100-strong academic research team, working on a range of research projects in the field of wireless multimedia communications sponsored by industry, the Engineering and Physical Sciences Research Council (EPSRC) UK, the European Research Council's Advanced Fellow Grant and the Royal Society's Wolfson Research Merit Award.  He is an enthusiastic supporter of industrial and academic liaison and he offers a range of industrial courses.  He is also a Governor of the IEEE VTS.  During 2008 - 2012 he was the Editor-in-Chief of the IEEE Press and a Chaired Professor also at Tsinghua University, Beijing.  His research is funded by the European Research Council's Senior Research Fellow Grant.  For further information on research in progress and associated publications please refer to http://www-mobile.ecs.soton.ac.uk Lajos has 18 000+ citations.

\end{IEEEbiography}


\begin{thebibliography}{11}

\bibitem{IEEEhowto:1}
Norton Corporation, ``The 2012 Norton cybercrime report," Sept.
2012, available on-line at
http://www.norton.com/2012cybercrimereport.

\bibitem{IEEEhowto:2}
M. E. Hellman, ``An overview of public key cryptography," \emph{IEEE
Commun. Mag.}, vol. 16, no. 6, pp. 42-49, May 2002.

\bibitem{IEEEhowto:3}
S. V. Kartalopoulos, ''A primer on cryptography in communications,"
\emph{IEEE Commun. Mag.}, vol. 20, no. 4, pp. 146-151, Apr. 2006.

\bibitem{IEEEhowto:4}
A. D. Wyner, ``The wire-tap channel," \emph{Bell System Technical
Journal}, vol. 54, no. 8, pp. 1355-1387, 1975.

\bibitem{IEEEhowto:5}
S. K. Leung-Yan-Cheong and M. E. Hellman, ``The Gaussian wiretap
channel," \emph{IEEE Trans. Inf. Theory}, vol. 24, pp. 451-456, Jul.
1978.

\bibitem{IEEEhowto:6}
S. K. Leung-Yan-Cheong, ``Multi-user and wiretap channels including
feedback," Ph.D. thesis, Stanford University, Stanford, CA, 1976.

\bibitem{IEEEhowto:7}
W. Chua, C. Yuen, Y. Guan, and F. Chin, ``Robust multi-antenna
multi-user precoding based on generalized multi-unitary
decomposition with partial CSI feedback," \emph{IEEE Trans. Veh.
Tech.}, vol. 62, no. 2, pp. 596-605, Feb. 2013.

\bibitem{IEEEhowto:8}
B. Aygun and A. Soysal, ``Capacity bounds on MIMO relay channel with
covariance feedback at the transmitters," \emph{IEEE Trans. Veh.
Tech.}, vol. 62, no. 5, pp. 2042-2051, May 2013.

\bibitem{IEEEhowto:9}
A. Khisti and G. Wornell, ``Secure transmission with multiple
antennas: The MISOME wiretap channel," \emph{IEEE Trans. Inf.
Theory}, vol. 56, no. 7, pp. 3088-3104, Jul. 2010.

\bibitem{IEEEhowto:10}
A. Khisti, G. Womell, A. Wiesel, and Y. Eldar, ``On the Gaussian
MIMO wiretap channel," in \emph{Proc. IEEE Int. Symp. Inf. Theory},
pp. 2471-2475, Nice, France, Jun. 2007.

\bibitem{IEEEhowto:11}
F. Oggier and B. Hassibi, ``The secrecy capacity of the MIMO wiretap
channel," \emph{IEEE Trans. Inf. Theory}, vol. 57, no. 8, pp.
4961-4972, Aug. 2011.

\bibitem{IEEEhowto:12}
A. Sendonaris, E. Erkip, and B. Aazhang, ``User cooperation
diversity - Part I: System description," \emph{IEEE Trans. Commun.},
vol. 51, no. 11, pp. 1927-1938, Nov. 2003.

\bibitem{IEEEhowto:13}
Y. Zou, Y.-D. Yao, and B. Zheng, ``Opportunistic distributed
space-time coding for decode-and-forward cooperation systems,"
\emph{IEEE Trans. Signal Process.}, vol. 60, no. 4, pp. 1766-1781,
Apr. 2012.

\bibitem{IEEEhowto:14}
M. Yuksel and E. Erkip, ``Secure communication with a relay helping
the wiretapper," in \emph{Proc. 2007 IEEE Information Theory
Workshop}, pp. 595-600, Lake Tahoe, CA, Sept. 2007.

\bibitem{IEEEhowto:15}
E. Tekin and A. Yener, ``The general Gaussian multiple access and
two-way wire-tap channels: Achievable rates and cooperative
jamming," \emph{IEEE Trans. Inf. Theory}, vol. 54, no. 6, pp.
2735-2751, Jun. 2008.

%\bibitem{IEEEhowto:16}
%G. Zheng, L.-C. Choo, and K.-K. Wong, ``Optimal cooperative jamming
%to enhance physical layer security using relays," \emph{IEEE Trans.
%Signal Process.}, vol. 59, no. 3, pp. 1317-1322, Mar. 2011.

\bibitem{IEEEhowto:16}
R. Zhang, L. Song, Z. Han, and B. Jiao, ``Physical layer security
for two-way untrusted relaying with friendly jammers," \emph{IEEE
Trans. Veh. Tech.}, vol. 61, no. 8, pp. 3693-3704, Aug. 2012.

\bibitem{IEEEhowto:17}
Y. Liang, H. V. Poor, and S. Shamai, ``Information theoretic
security", MA: Now Publishers, vol. 5, no. 4-5, pp. 355-580, 2008.

\bibitem{IEEEhowto:18}
Z. Ding, M. Xu, J. Lu, and F. Liu, ``Improving wireless security for
bidirectional communication scenarios," \emph{IEEE Trans. Veh.
Tech.}, vol. 61, no. 6, pp. 2842-2848, Jun. 2012.

\bibitem{IEEEhowto:19}
A. O. Hero, ``Secure space-time communication," \emph{IEEE Trans.
Inform. Theory}, vol. 49, no. 12, pp. 3235-3249, Dec. 2003.

\bibitem{IEEEhowto:20}
C. E. Shannon, \textquotedblleft A mathmatical theory of
communication," \emph{Bell System Technical Journal}, vol. 27, pp.
379-423, 1948.

\bibitem{IEEEhowto:21}
{R. Yin, \emph{et al.}, ''Tradeoff between reliability and security in block ciphering systems with physical channel errors," in \emph{Proc. 2010 IEEE Military Communications Conference (MILCOM 2010)}, San Jose, CA, Oct. 2010.}

\bibitem{IEEEhowto:22}
Y. Zou, Y.-D. Yao, and B. Zheng, ``An adaptive cooperation diversity
scheme with best-relay selection in cognitive radio networks,"
\emph{IEEE Trans. Signal Process.}, vol. 58, no. 10, pp. 5438-5445,
Oct. 2010.

\bibitem{IEEEhowto:23}
X. Wang, K. Wang, and X.-D. Zhang, ``Secure relay beamforming with
imperfect channel side information," \emph{IEEE Trans. Veh. Tech.},
vol. 62, no. 5, pp. 2140-2155, May 2013.


\end{thebibliography}
\end{document}